\documentclass[pdflatex,sn-nature]{sn-jnl}

\usepackage{graphicx}%
\usepackage{multirow}%
\usepackage{amsmath,amssymb,amsfonts}%
\usepackage{amsthm}%
\usepackage{mathrsfs}%
\usepackage[title]{appendix}%
\usepackage{xcolor}%
\usepackage{textcomp}%
\usepackage{manyfoot}%
\usepackage{booktabs}%
\usepackage{algorithm}%
\usepackage{algorithmicx}%
\usepackage{algpseudocode}%
\usepackage{listings}%
\usepackage{url}%

\theoremstyle{thmstyleone}%
\theoremstyle{thmstyletwo}%

\theoremstyle{thmstylethree}%

\begin{document}

\title[Heterogeneity of household stock portfolios ]{Heterogeneity of household stock portfolios in a national market}


\author[1]{\fnm{Matteo} \sur{Milazzo}}

\author[2]{\fnm{Federico} \sur{Musciotto}}

\author[3]{\fnm{Jyrki} \sur{Piilo}}

\author*[2,4]{\fnm{Rosario N.} \sur{Mantegna}}\email{rosario.mantegna@unipa.it}

\affil[1]{\orgdiv{Department of Physics and Astronomy “Ettore Majorana”}, \orgname{University of Catania}, \orgaddress{\street{Via S. Sofia, 64}, \city{Catania}, \postcode{95123}, \country{Italy}}}

\affil[2]{\orgdiv{Department of Physics and Chemistry ``Emilio Segrè"}, \orgname{University of Palermo}, \orgaddress{\street{Viale delle Scienze, Edificio 18}, \city{Palermo}, \postcode{90128}, \country{Italy}}}

\affil[3]{\orgdiv{Department of Physics and Astronomy}, \orgname{University of Turku}, \orgaddress{\street{Turun Yliopisto}, \city{Turku}, \postcode{20014}, \country{Finland}}}

\affil[4]{\orgname{Complexity Science Hub Vienna}, \orgaddress{\street{Metternichgasse 8}, \city{Vienna}, \postcode{1030}, \country{Austria}}}


\abstract{
We study the long term dynamics of the stock portfolios owned by single Finnish legal entities in the Helsinki venue of the Nasdaq Nordic between 2001 and 2021. Using the Herfindahl-Hirschman index as a measure of concentration for the composition of stock portfolios, we investigate the concentration of Finnish household portfolios both at the level of each individual household and tracking the time evolution of an aggregated Finnish household portfolio. We also consider aggregated portfolios of two other macro categories of investors one comprising Finnish institutional investors and the other comprising foreign investors. Different macro categories of investors present a different degree of concentration of aggregated stock portfolios with highest concentration observed for foreign investors. For individual Finnish retail investors, portfolio concentration estimated by the Herfindahl-Hirschman index presents high values for more than half of the total number of retail investors. In spite of the observation that retail stock portfolios are often composed by just a few stocks, the concentration of the aggregated stock portfolio for Finnish retail investors has a portfolio concentration comparable with the one of Finnish institutional investors. Within retail investors, stock portfolios of women present a similar pattern of portfolios of men but with a systematic higher level of concentration observed for women both at individual and at aggregated level.
}

\keywords{Financial market, market heterogeneity, market participants, market ecology, Nasdaq Nordic market, domestic and foreign investors.}



\maketitle

\section{Introduction}\label{sec1}

A line of academic and professional research considers financial markets as complex systems \cite{mantegna1999introduction, bouchaud2003theory,sornette2009stock,farmer2024making}. 
Financial markets are social and economic infrastructures pursuing different goals. 
Several financial markets exist and each of them is specialised in the trading of one type, or few types, of financial assets. In this study we focus on the stock market that is one of the most prominent financial market. 
Starting from the eighties of the last century a progressive stock market liberalisation has diffused in mature and emerging markets. Thanks to this process of liberalization, i.e., a decision by a country's government to allow foreigners to purchase shares in that country's stock market, today foreign and domestic investment  is possible and quite easily performed thanks to broker intermediation in the majority of world stock markets. 

Stock markets provide credit and investment opportunities to companies and individuals. Market participants in the stock market are therefore inherently of different type. In fact we observe households, e.g., retail investors, companies (both financial and non financial companies) and various types of institutions. These investors can be domestic or foreign with respect to the country hosting the market venue.   

Classic financial theory predicts that this type of heterogeneity can be subsumed into an appropriate rational agent operating in the market \cite{hommes2006heterogeneous,kirman2010complex}. The study of idiosyncratic behaviours associated with the different categories of market participants is a main area of research of behavioural finance \cite{barberis2003survey}.  Financial market ecology has also been investigated in the literature  \cite{farmer1999frontiers,farmer2002market,lo2004adaptive,preda2019noise}. 

Performing empirical studies of behavioural finance on a scale covering an entire country is extremely difficult if not impossible. The reason of the difficulty lies on two facts. Firstly, investment data at the level of the single investor are data to be kept confidential to protect the privacy of the investor. Secondly, in most of western countries the process of clearing and settlement of stocks is in the majority of cases performed through nominee  registration. Nominee registration means that a legal entity, a nominee, is registered in a register of owners of financial instruments instead of the beneficial owner. With this recording mechanism distributed across many chains of custodians, it is extremely difficult to have access to financial ownership of all legal entities operating in a country for all institutions excluding law enforcement agencies and supervisory agencies.
In the Western world it exists only an exception to this standard way of tracking stock ownership. This is the case of Finland and, for this reason, stock ownership in Finland is the focus of the present study. 
In fact financial ownership of Finnish legal entities have been investigated by many studies since 2000 \cite{grinblatt2000investment, grinblatt2001distance, grinblatt2009sensation, tumminello2012identification, lillo2015news, musciotto2016patterns, musciotto2018long,  baltakiene2019clusters}. Some of these studies have investigated investment pattern of Finnish investors over very long period of time \cite{musciotto2021high, keloharju2021quarter}. Having the balance at the start of the period and all the transactions for each asset traded by each legal entity, we were able to reconstruct the portfolios for each one of them. In this work we focused on the analysis of portfolios choices made, in particular, by households.

In this paper we consider three macro categories of market participants:
(i) retail investors, i.e., individual investors. In the technical literature they are also called households; (ii) non-retail investors, which include institutional investors like banks, insurance companies, investment firms and funds, but also non financial companies operating in various industries, national and local governmental organizations and non-profit institutions; (iii) foreign investors both as nominee registered investors or as foreign legal entity directly investing into stocks. 

Among these macro category of investors, Finnish households, i.e. retail investors, are the largest number of market participants in the Helsinki venue of the Nasdaq Nordic stock exchange. Here we present an empirical study on the long term evolution of household investment in stocks performed at the level of an entire country.

Price discovery in the market is emerging due to the investment decisions of all market participants. In a long term study covering more than 20 years, there are many aspects to take into account to properly analyze the system. The number of investors changes over time and the same is true for the number of stocks listed in the stock exchange. Each listed company can change the number of issued stocks, and they can also merge or split into new entities. Stock splits (direct and inverse) are frequently observed. Another aspect influencing the evolution of the market can be represented by regulatory changes occurring at the level of trading , clearing or settlement. Change on the society as, for example, the emergence of coordination through social networks can also affect the way price discovery occurs in a stock market \cite{welch2022wisdom}.

With the present study, we aim to answer to the following questions: (i) How did market participants and in particular household investors stock  portfolios change over a period of time
of more than 20 years? (ii) Is it possible to characterize quantitatively how the heterogeneity of the system evolve in time? In the present study we will investigate empirically the degree of stock portfolio concentration and the wealth inequality in stock portfolio investment for all legal Finnish entities and for all foreign investors investing in stocks in the Helsinki venue of the Nasdaq Nordic market over a period of 21 years. The portfolio concentration will be quantified through the Herfindahl-Hirschman Index (HHI) \cite{hirschman1964paternity, hall1967measures} and the inequality of stock investment will be quantified through the Lorenz curve \cite{gastwirth1972estimation}. 

The dataset we investigate is provided by the Finnish clearing house Euroclear Finland. It includes all the clearing and settlement records related to Finnish stocks traded at the Helsinki venue of the Nasdaq Nordic market by each investor. In the present study, a single investor is a specific legal entity, it can therefore be a household, a company or a governmental organization. We believe this is the best way to characterize single investors because any ownership of stock is directly related to a specific legal entity. The data include also some metadata for  some legal entities such as birth year, gender and postal code when applicable. We focus on the time period elapsing between April 2001 and December 2021. This choice is motivated by the availability of historical information about the OMX Helsinki all-shares index. In fact, historical information about stocks traded in a market are difficult to obtain over long time period due to the bias on updated information present in the services of financial providers. In our study the most reliable information we were able to find concerned the period from 2001 to 2021 and therefore we used it for our study.

The paper is organized as follows. In Section \ref{sec2} we discuss the structure and properties of the investigated dataset. In Section \ref{sec3} we describe the Herfindahl-Hirschman Index whereas in Section \ref{sec4} we present our empirical results discussing the concentration of stock portfolios of Finnish households, the inequality pattern of the households investment and the concentration properties and time evolution of aggregated portfolios of macro categories of investors. In Section \ref{sec5} we draw some conclusions.   

\section{A unique dataset}\label{sec2}

The main source of the data used in this paper is the ownership data of financial assets issued in Finland collected by the clearing house Euroclear Finland. Euroclear Finland performs all actions needed for the issuance settlement and asset services occurring in Finland or referring to Finnish financial assets. As a clearing and settlement house, it verifies and records the changes in holding of all the financial assets issued in Finland on a daily timebase. These financial assets include the stocks traded at the Helsinki venue of the Nasdaq Nordic Market. The dataset currently collected by Euroclear Finland and formerly collected by the Finnish Central Securities Depository (FCSD). The register has the shareholdings in Finnish issued stocks of all Finnish investors and of all foreign investors asking to
exercise their vote right, both retail and institutional. The register also includes information concerning the ownership of stocks for foreign investors that are nominee registered. Nominee registration means that a legal entity, a nominee, is registered in a register of owners of financial instruments instead of the beneficial owner. The database records official ownership of financial assets and the trading records are updated on a daily basis accordingly. The records include all the transactions, executed in worldwide stock exchanges and in other venues, which change the ownership of the assets. The database classifies investors into six main categories: non-financial corporations,
financial and insurance corporations, general governmental organizations, non-profit
institutions, households and foreign organizations. The database has been collected since
1 January 1995. We have access to the database for the period 1995–2021, under a special
agreement with Euroclear Finland. 

In the dataset, each legal entity has a shadowed unique ID and several other metadata such as gender, birth year and postal code. Each item of the dataset describes a change in ownership of a specific asset identified by its unique International Securities Identification Number (ISIN). Each record includes the date in which the change was registered, the number of shares involved and their price.

It is worth to spend some words to appreciate the uniqueness of this dataset. Finland is the only European country and one of the few countries of the world requiring the registration of the ownership of financial assets at the level of each Finnish legal entity (households, firms, public organization, etc.). Essentially in all Western countries, legal entities uses indirect registration, i.e., so-called registration achieved as nominee registered (see for details \url{https://www.europeanissuers.eu/positions/files/view/591da562f05ba-en}). 

In other words, financial asset ownership data are not only hard to access due to confidentiality reasons, but they are hardly covering a large set of population because their custody is usually performed by using the so-called method of nominee registering. It implies that the ownership of each financial asset is registered under a custodian, i.e., a bank or a clearing house, and therefore it is not possible to access comprehensive records of the trading activity performed by different legal entities belonging to an entire country. The only exception to this general policy that is present in Western countries is essentially Finland. In fact, Finland is the only country of the Western world where the ownership of financial assets issued in Finland are tracked on a daily level for each Finnish legal entity due to a legal requirement. For this uniqueness, Finland constitutes a “laboratory” for the investigation of the ecology of market participants acting on multiple time scales in a financial market.

Having access to the complete history of investment decisions about Finnish financial assets of single individuals and, more generally, of any Finnish legal entity, allows to study in great detail the long term dynamics of a financial market with respect to the trading decision of the legal entities of the same country. In recent years many studies have been focused on the investment decision of Finnish legal entities. 

Here we focus on heterogeneity of stock portfolio of a Finnish legal entity investing into Finnish stocks. In fact, having access to the historical record of the financial ownership of all Finnish legal entities from 2001 to 2021 we can reconstruct the time evolution of stock portfolio of each Finnish legal entity over time. 

Financial assets issued in Finland includes stocks, bonds and derivatives. In the present study we focus our attention on stock portfolios. Our choice is to consider all stocks that are included in the OMX Helsinki all-shares (OMXHPI) i.e., the index including all the stocks traded in the Helsinki venue of the Nasdaq Nordic, including small, mid and large-cap stocks. The number of stocks changes over time. In fact there are new companies entering and companies exiting the market or merging with other companies. Each company can also have more than one stock, merge or split them into different assets. Our study covers a time period of more than 20 years on a daily scale. On this long time period, we believe a reasonable timescale properly describing portfolio rebalancing is a monthly timescale. Therefore in the present study, for each legal entity,  we are considering stock portfolios owned at the last day of each month. We complement the information present in the Euroclear database with financial information present in the Eikon database from Refinitiv  (\url{https://eikon.refinitiv.com}). Eikon was primarily used to track back the changes of the index. Information stored in Eikon was compared with information obtained from the Nasdaq Press Center (\url{https://www.nasdaqomxnordic.com/news/press_center)}. Eikon information about OMXHPI dates back to 2001. For this reason we are analysing data during the time period 2001-2021. During this time period the number of stocks composing the index ranged from 129 to 157. As supplementary information we provide the complete list of financial assets included into the OMXHPI index during the period from 2001 to 2022. The information includes the estimated initial and final dates of stock inclusion. 

Our set of stocks includes all stocks belonging to the OMXHPI index with the only exception of stocks of telecommunication company Elisa. The reason we remove this specific stock is due to the peculiar policy this company has followed about its shares. 
Elisa is one of the major telecommunication Finnish companies
and during 1999 it gifted every holder of a membership contract 150
shares of the company. Customers of Elisa receiving Elisa's shares were 
several hundred thousands households meaning that a huge amount of legal entities
became owners of some shares of the company by the decision of the company and not by
a decision of the investors. Most of these household never traded the stock 
at all or for several years.
Due to the peculiar policy followed by this company we decided to remove this stock from
our analysis. For information about Elisa share history visit the web site 
\url{https://elisa.com/corporate/investors/share/share-history/}.

\section{Heterogeneity measure}\label{sec3}
In this study we are primarily interested into the heterogeneity of stock portfolios of Finnish legal entities with a special focus on households, i.e. retail, investors. We therefore will systematically use a measure of concentration for owned stock portfolios.  

Specifically, we quantify portfolio concentration, with one of the most popular measures of concentration that is the Herfindahl-Hirschman Index (HHI) \cite{hirschman1964paternity, hall1967measures}, defined as follows:
\begin{equation}
	H=\sum_{i=1}^{N}{{w_i}^2}
\end{equation}
where $w_i$ is the weight, i.e., the fraction of value of the $i$-th stock present in the portfolio of size $N$ and $\sum_{i=1}^{N}w_i=1$. The HHI ranges from $1/N$ (when the same weight is given to each stock) to $1$ (when portfolio has only one stock).

In this study, for each calendar month the HHI is calculated both for the stock portfolio of each Finnish legal entity owning a stock portfolio and for aggregated portfolios representing the ownership of financial stocks of some macro category of legal entities. The way we compute aggregated portfolios is by considering all the holdings of stocks owned by a specific macro category at a given day. In our analysis, as macro categories we choose(i) households,  (ii) non-retail investors, and (iii)  foreign investors. We will also consider subsets of the macro categories as the ones obtained by considering women and men investors separately. Non-retail legal entities include non-financial corporations, financial and insurance corporations, general governmental organizations, and non-profit.
Foreign investors include nominee registered investors and foreign registered legal entities.  

Both for individual legal entities and for aggregated macro categories, stock portfolios are computed at the last day of each calendar month from 30/04/2001 to 31/12/2021, a period covering more than 20 years.

\section{Results}\label{sec4}

\subsection{HHI distribution of household portfolios} 
The first observation concerns the level of investment concentration in household portfolios. As discussed above we use the HHI as a measure of portfolio concentration. According with previous results obtained in the literature of behavioral finance, we observe that stock portfolios of households present a degree of concentration that is not compatible with basic models of investment diversification and risk management of theoretical finance.

\begin{figure}[h]
\centering
        \includegraphics[width=0.7\textwidth]{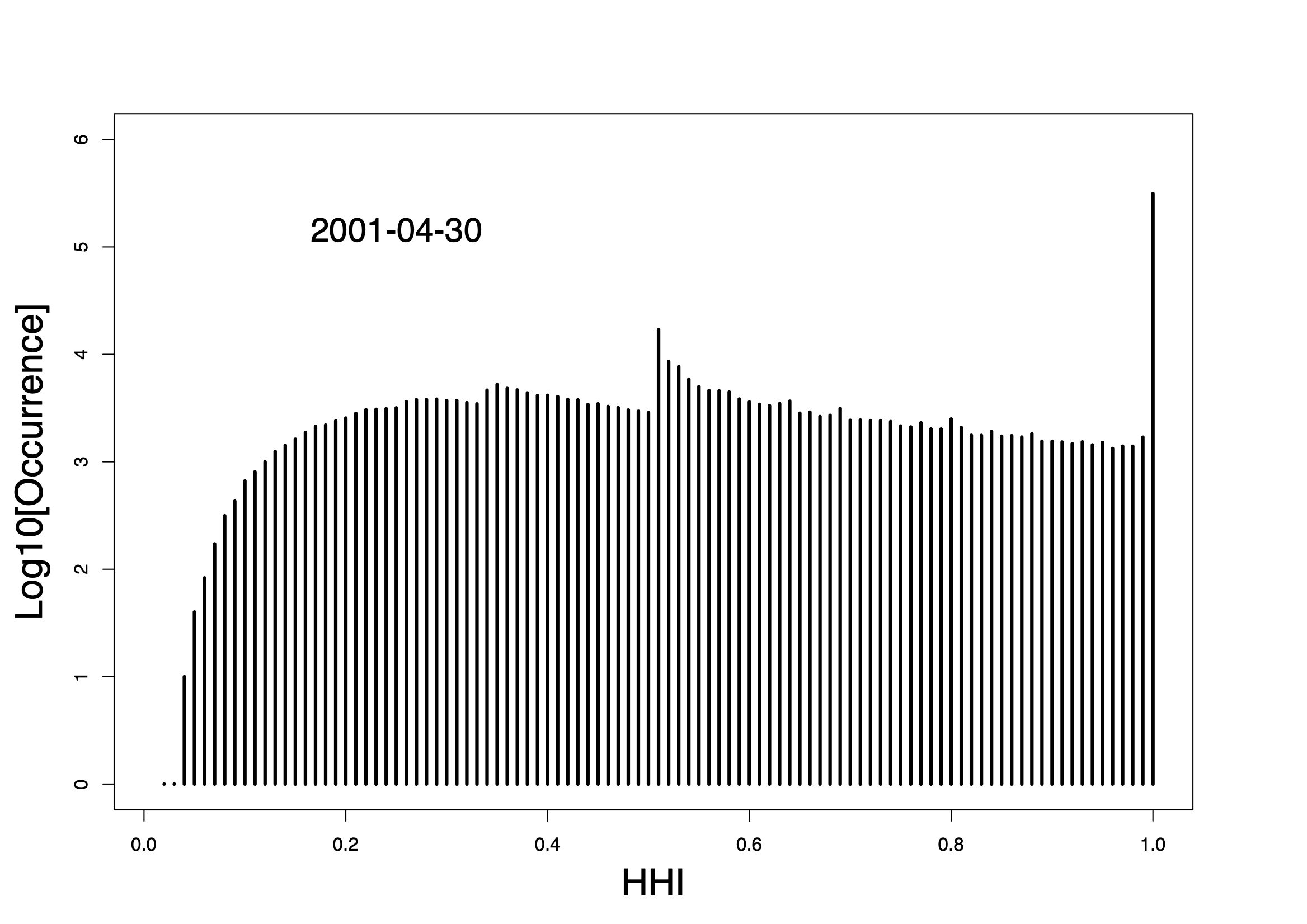}
        \includegraphics[width=0.7\textwidth]{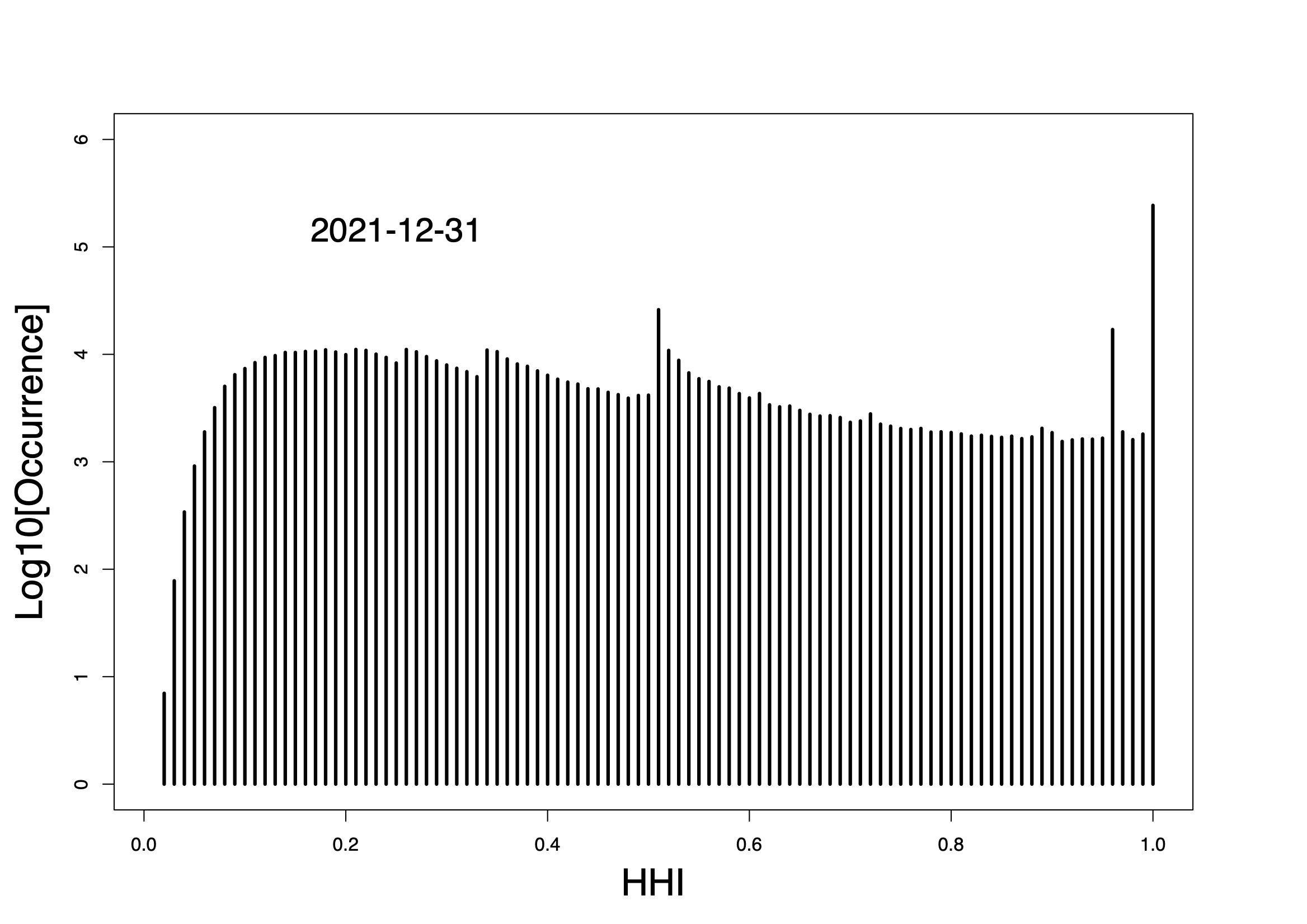}
        \caption{Occurrence of HHI values of household portfolios. Household portfolios  at April 30, 2001  (top panel), total number of portfolios is 592,465.  Household portfolios  at December 31, 2021 (bottom panel), total number of  household portfolios is 776621. Note that the y-axis is in a base 10 logarithmic scale.}
        \label{fig:pdf_hhi}
\end{figure}

Our study is covering a long period of time and therefore we are able to track the long term changes in the concentration attitude of Finnish household. To this end we first consider the pattern of the histogram of occurrence of HHI values in household portfolios. In Fig.  \ref{fig:pdf_hhi} we show the histogram  of the number of Finnish households having a given HHI value for the months of April 2001 (top panel) and December 2021 (bottom panel).  Fig.  \ref{fig:pdf_hhi} allows to analyze distribution of HHI values characterizing portfolios of Finnish household both for the first and for the last month of the time period of our investigation. Both histograms show that HHI observed values are ranging from approximately $1/N$ to $1$, where $N$ is the number of stocks traded at Helsinki venue of Nasdaq Nordic. Considering that $N$ values were $156$ in $2001$ and $143$ in $2021$, the expected HHI minimum value is $HHI_{min}=1/156=6.41 \times 10^{-3}$ for 2001 and $HHI_{min}=1/143=6.99 \times 10^{-3}$ for 2021.
In both months, household portfolios shown a high level of heterogeneity with HHI values covering the entire interval from $HHI_{min}$ to one in an approximately uniform way. In addition to this approximately uniform distribution we notice a few prominent peaks. The most relevant one is the peak observed for $HHI=1$, namely the number of household portfolios with just a single stock. This number is pretty high in all months and covers a large fraction of all household portfolios. Specifically, its occurrence is 313651 in 4/2001 (top panel of Fig.  \ref{fig:pdf_hhi}) when the total number of household portfolios is 592465 and its occurrence is 243481 in 12/2021 (bottom panel of Fig.  \ref{fig:pdf_hhi}) when the total number of household portfolios is 776621. In addition to this prominent peak, other peaks are observed for HHI values equal to 0.5, 1/3, 0.25 indicating equal weight portfolios composed by 2, 3, or 4 stocks respectively. In 2021 there is also a distinct peak at $HHI \approx 0.96$ primarily due to household portfolios of two stocks with a weight ratio of about 40. 

Over the 20 year period the profile of the histogram has not changed too much. However, some change is evident, The fraction of household portfolios with just a single stock has diminished from $52.9 \%$ to $31.3 \%$. Another prominent change involves occurrences for lowest values of the HHI. From the figure is evident that the occurrence of HHI values below 0.2 increases both in absolute and in relative terms. 

\begin{figure}[h]
\centering
\includegraphics[width=0.9\textwidth]{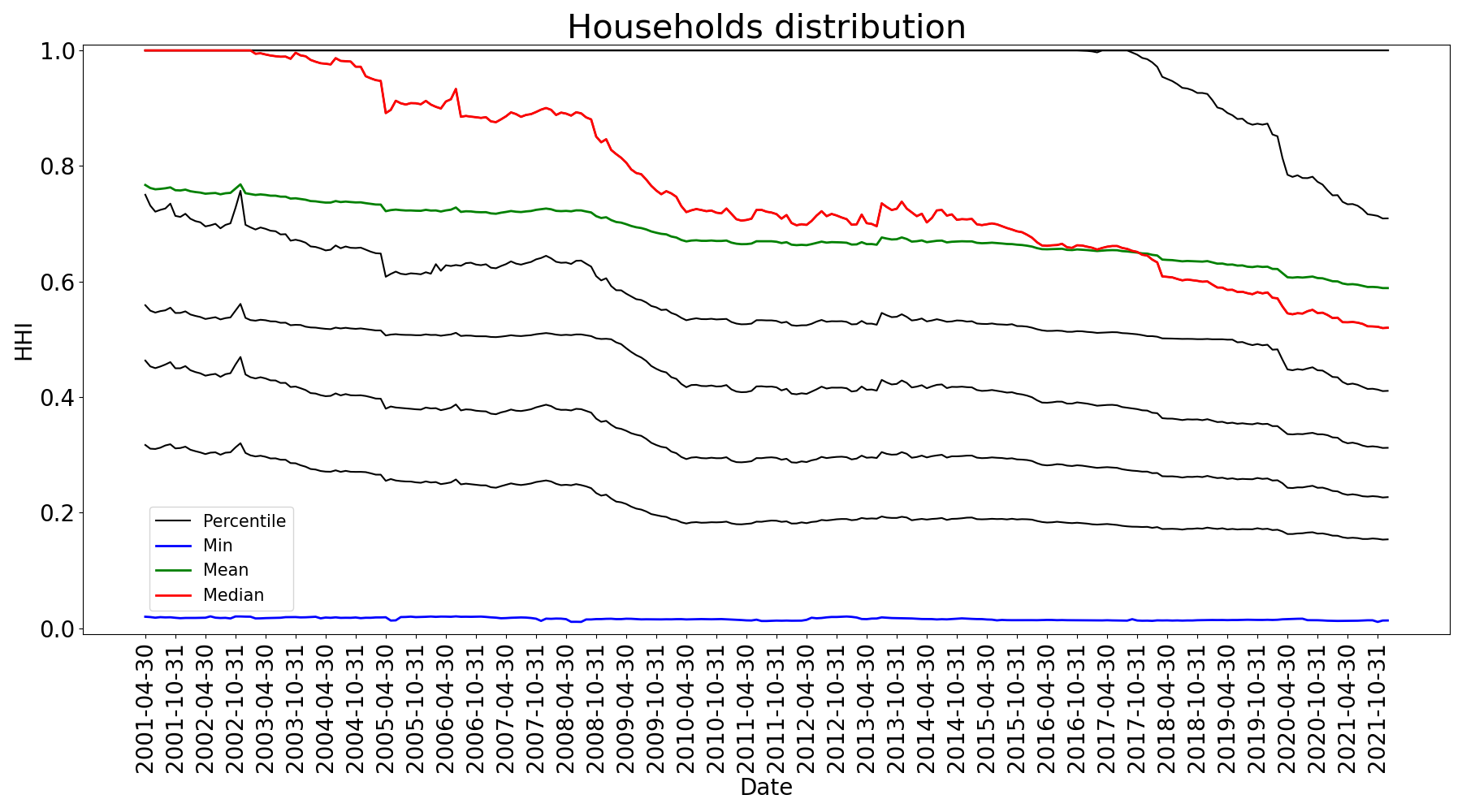}
\caption{Time evolution of HHI deciles of HHI distribution (black lines). The deciles start from bottom to top. The red line is the median of HHI values while green line is the mean. Both the median and the mean show a decreasing dynamics over the years. Note that still on December 2021 more than $30\%$ of household portfolios have HHI=$1$ (i.e., they are composed by just a single stock).
}\label{fig:hhi_dec}
\end{figure}

The analysis of the two panels of Fig. \ref{fig:pdf_hhi} therefore suggests the presence of some trend in the degree of concentration of household portfolios. To quantitatively evaluate this trend we compute household portfolios monthly and we plot the mean value of HHI calculated for each household investor. The result of this analysis is shown in Fig. \ref{fig:hhi_dec}.
In the figure we note a slow dynamics in the mean value of HHI for household portfolios. In fact the mean value decreases from a value close to 0.8 to a value close to 0.6 in about 20 years. The decrease is stronger in the median that is decreasing from 1 to less than 0.6.  Also the first decile (lowest black line) is decreasing especially in the time period from 2001 to 2009.
In summary, Fig. \ref{fig:hhi_dec} clearly shows that the degree of concentration of household portfolios has been slowly changing during the last 20 years with a neat tendency to make household portfolios less concentrated.
The mean and median values of HHI remains still very high and quite far from what is expected by financial theory but a slow dynamics of portfolio concentration is certainly present progressively moving towards values indicating less concentration.

\subsection{Inequality of household portfolios} 
\label{inequality}
Heterogeneity is present in several dimensions of household portfolios. In the previous subsection we have considered the degree of concentration of stock portfolios, here we consider the distribution of the value of stock portfolios. The value of stock portfolios for households spans many order of magnitude. For example on 2021-12-13 the interval between the lowest and the highest value of stock portfolio is spanning more than 10 orders of magnitude. To evaluate the degree of heterogeneity of the value of the portfolios we use the classic Lorenz curve.   In a Lorenz curve the cumulative percentage of of a given amount (the classic example is national income) is plotted against the cumulative percentage of the corresponding population (ranked in increasing size of share). The extent to which the curve distantiates from below a straight diagonal line indicates the degree of inequality of distribution.  
\begin{figure}[h]
\centering
\includegraphics[width=0.9\textwidth]{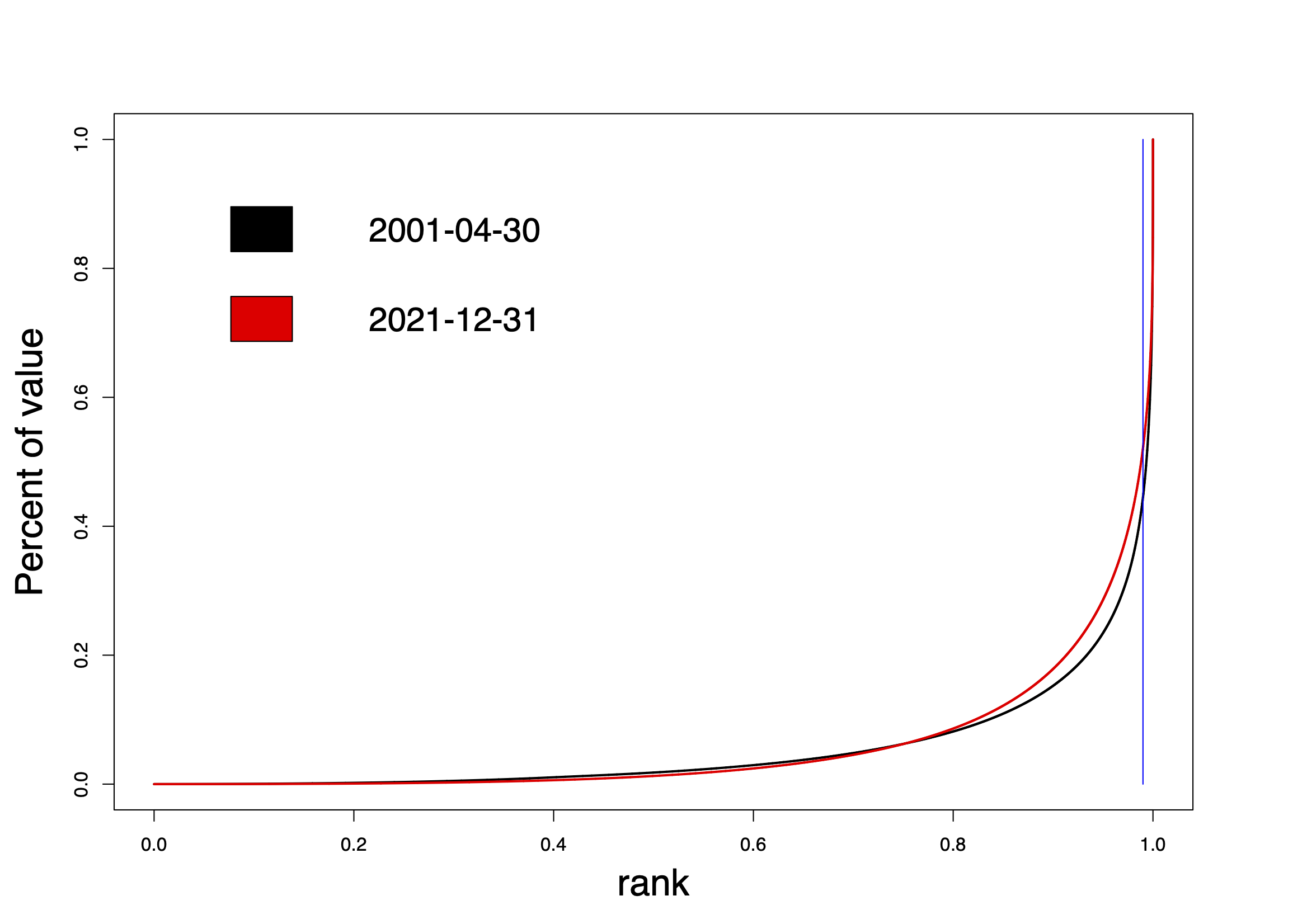}
\caption{Lorenz curve of the cumulative value of household stock portfolios for the day 2001-04-30 (black line) and for the day 2021-12-31 (red line). The vertical line highlights the wealthier $1\%$ investors. These investors own $55.5\%$ or $47.9\%$ of the total wealth for 2001-04-30 and 2021-12-31 respectively. 
}\label{fig:lorenz_hou}
\end{figure}
In Fig. \ref{fig:lorenz_hou} we show the Lorenz curve observed for household stock portfolios computed for the day 2001-04-30 (black line) and for the day 2021-12-31 (red line). These two examples are representative of the entire period of time.  In both cases, Lorenz curves strongly deviate from the diagonal indicating a pronounced level of inequality in the value of household portfolios. In fact, the curves show that wealthier $1\%$ investors (pointed out by the vertical blue line) own $55.5\%$ or $47.9\%$ of the total wealth for 2001-04-30 and 2021-12-31 respectively. The number of these investors is 5924 and 7766 households respectively. The two curves essentially overlaps for low values of the rank and for high values of the percent of value. However, in the middle part of the curve we observe a difference between the cases with a slightly diminished inequality present in 2021.
The analysis of the Lorenz curve shows that although the number of households could be very large (in the present analysis is larger than 500,000 households, about half of the total wealth allocation of the aggregated household portfolio is decided by a number of investors of the order of several thousands.

\subsection{HHI of macro categories aggregated portfolios} 
\label{macrocategories}
In this subsection, we consider the HHI of portfolios owned by a few macro categories of investors. The macro categories we consider are the following: (a) Households, (b) Institutional investors, and (c) Foreign investors. 
In the {\it Households} macro category we group all Finnish households. In the macro category {\it Institutional investors} we group all Finnish legal entities belonging to (i) non financial corporations, (ii) financial and insurance corporations, (iii) governmental institutions, and (iv) non profit organizations. The last macro category is {\it Foreign investors}. It groups together nominee registered investors (under Finnish law only foreigners can be nominee registered) and foreigners registered in Finland as owners of Finnish assets. For reference, we also compute what we call the {\it market portfolio}, i.e., a portfolio aggregating all the ownership of stocks in a single portfolio.
\begin{figure}[h]
\centering
\includegraphics[width=0.9\textwidth]{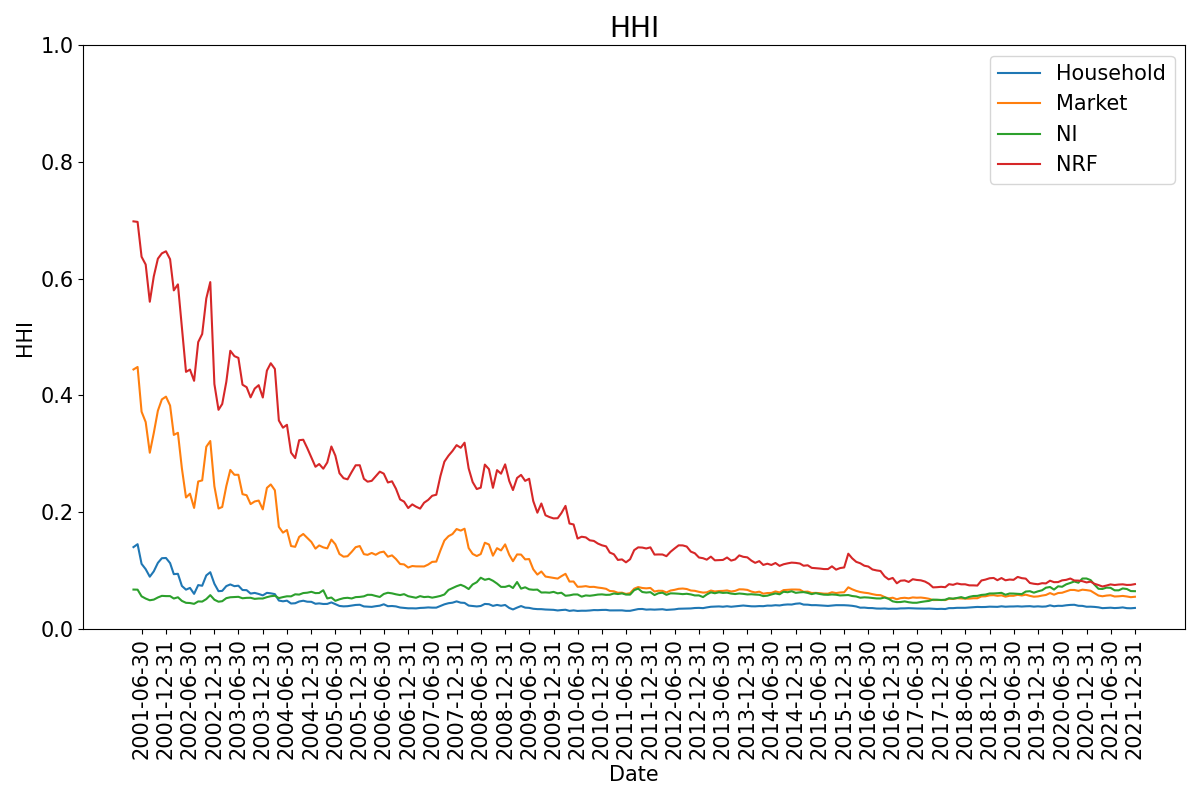}
\caption{HHI as a function of time for aggregated portfolios of (a) Finnish households (blue line), (b) Finnish institutional investors (green line), and (c) foreign investors (red line). We also provide the HHI for the whole market for comparison (orange line). }\label{fig:HHI_agg}
\end{figure}
In Fig. \ref{fig:HHI_agg} we show the temporal evolution of the HHI for the three macro categories defined above and for the entire market (orange line). The figure shows that foreign investors (red line) are characterized by an aggregated portfolio that is more concentrated than the aggregated portfolio of the entire market.
Portfolio concentration of aggregated Finnish investors is always less than the concentration of portfolio of foreign investors. 

The figure shows two distinct periods. One observed between 2001 and 2011 and the other observed between 2011 and 2021. During the first period the concentration of the market portfolio is pretty high with a HHI starting from about 0.4 and non monotonically declining to a value of about 0.1. Starting from 2011 changes in the HHI value become limited and the HHI index present an almost constant time evolution.  A similar pattern is observed for the aggregated portfolio of foreign investors. Indeed, for this macro category the degree of concentration is even higher than the one of the entire market especially in the time period from 2001 to 2011. We interpret such a high level of concentration of foreign investment into Finnish stock as primarily due to the prominent role that the Finnish company Nokia played internationally during and just after the so-called dotcom bubble. In fact, on 2001 technological stocks were at their highest quotation and Nokia stock price reached its maximum value and attracted a huge number of investors worldwide. Due to the fact that the majority of these investors are nominee registered, unfortunately, we are not able to discriminate among these foreign investors the fraction of investors that are retail. Most probably, a large fraction of them could be of retail type but assessing the precise fraction of it is impossible.  After the bursting of the dotcom bubble several nominee registered investors disinvested Nokia and other technological stocks and therefore portfolio concentration of foreign investors progressively diminished.   

In Fig. \ref{fig:HHI_agg} it is worth noting the pronounced difference of portfolio concentration between aggregated foreign and Finnish portfolios. In fact the aggregated portfolios of Finnish households (blue line in the figure) and institutions (green line in the figure) have values of the HHI indicating a much lower portfolio concentration. This is true for both periods discussed above but the discrepancy is extremely pronounced in the first period and especially just after the dotcom bubble (i.e., in 2001-2003). In other words, as aggregated investors, Finnish investors were able to disinvest Nokia and technological stocks much earlier than foreign investors participating to the technological stocks of the Nasdaq Nordic market. 
\begin{figure}[h]
\centering
\includegraphics[width=0.9\textwidth]{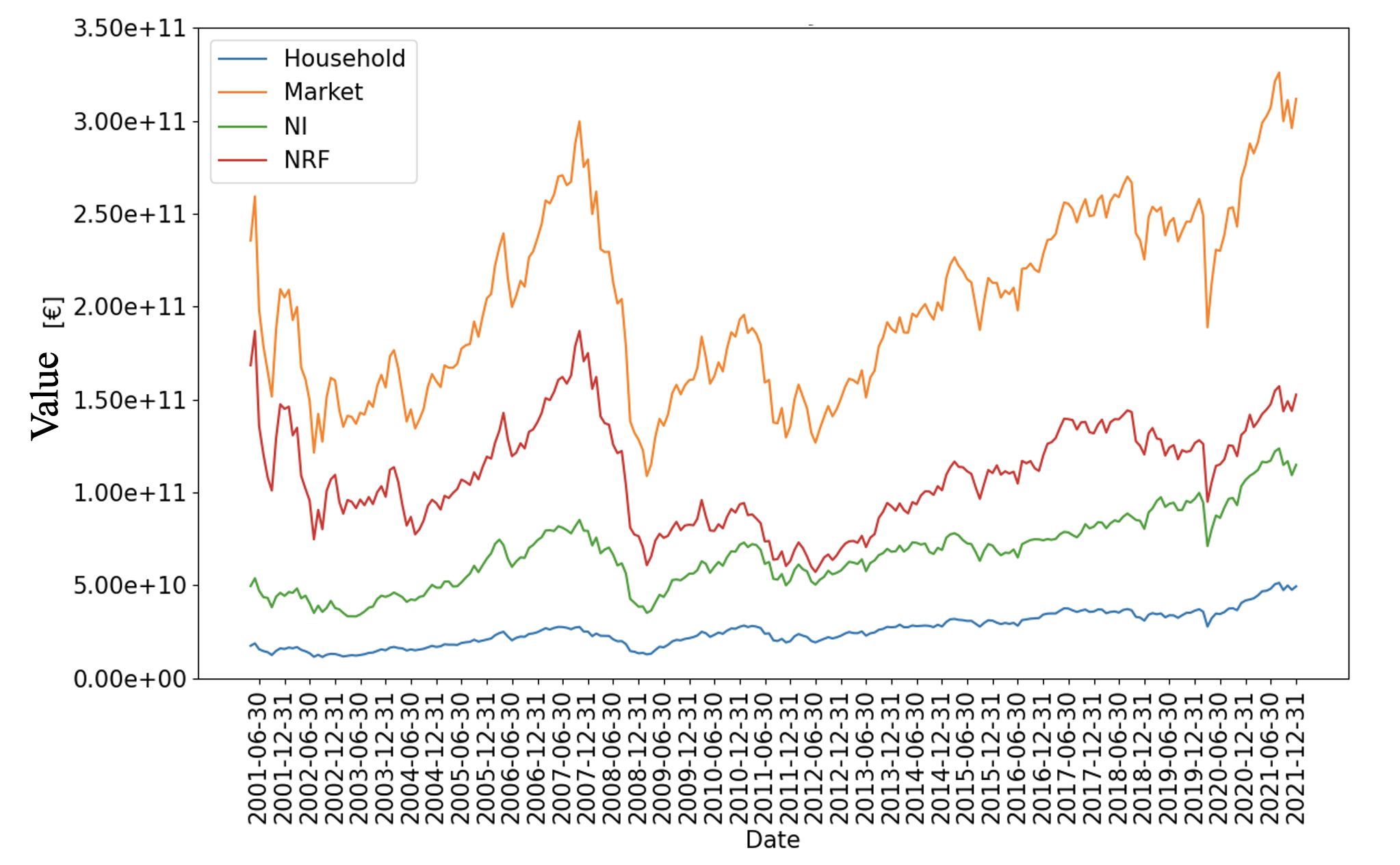}
\caption{Time evolution of the value in euro of aggregated portfolios of different macro categories. Specifically, we have Finnish households (blue line), Finnish institutional investors (green line), and foreign investors (red line). We also provide the capitalization for the whole market for reference (orange line). }\label{fig:money}
\end{figure}

In Fig. \ref{fig:money}  we show the total money invested by macro category as a function of time to better appraise the impact of the three macro categories investing in the market . Fluctuations in the capitalization value are the result of the price evolution of each individual stock and of the entering in and exiting from the market of some stocks. The global trend is the same for all the macro categories with an overall increase of market capitalization for long period of times with exceptions in 2008 and 2020 when two major market crises took place. In the Fig. \ref{fig:money} it is evident that foreign investors own the largest fraction of the market during the considered period. The second largest fraction is owned by Finnish institutional investors whereas households investors own the smallest share. It is rather common in several markets that  besides being huge in number, the impact in terms of global ownership of the retail investors is lower than the one of institutional and foreign investors.

\subsection{Wisdom of the crowd and gender difference} 
In this subsection, we consider household portfolio choices from the perspective of the so-called  \textit{wisdom of the crowd} \cite{surowiecki2005wisdom}. In other words, we will discuss on how a set of concentrated portfolios result in a much less concentrated aggregated portfolio. The majority of household investors do not follow canonical financial theory ending up into rather idiosyncratic stock allocation. In spite of this idiosyncratic approach not consistent with best known risk control practices, the aggregated market portfolios turns out to be compatible to a way lower level of stock concentration, showing that the whole category is collectively more financially wise.

\begin{figure}[h]
\centering
\includegraphics[width=0.9\textwidth]{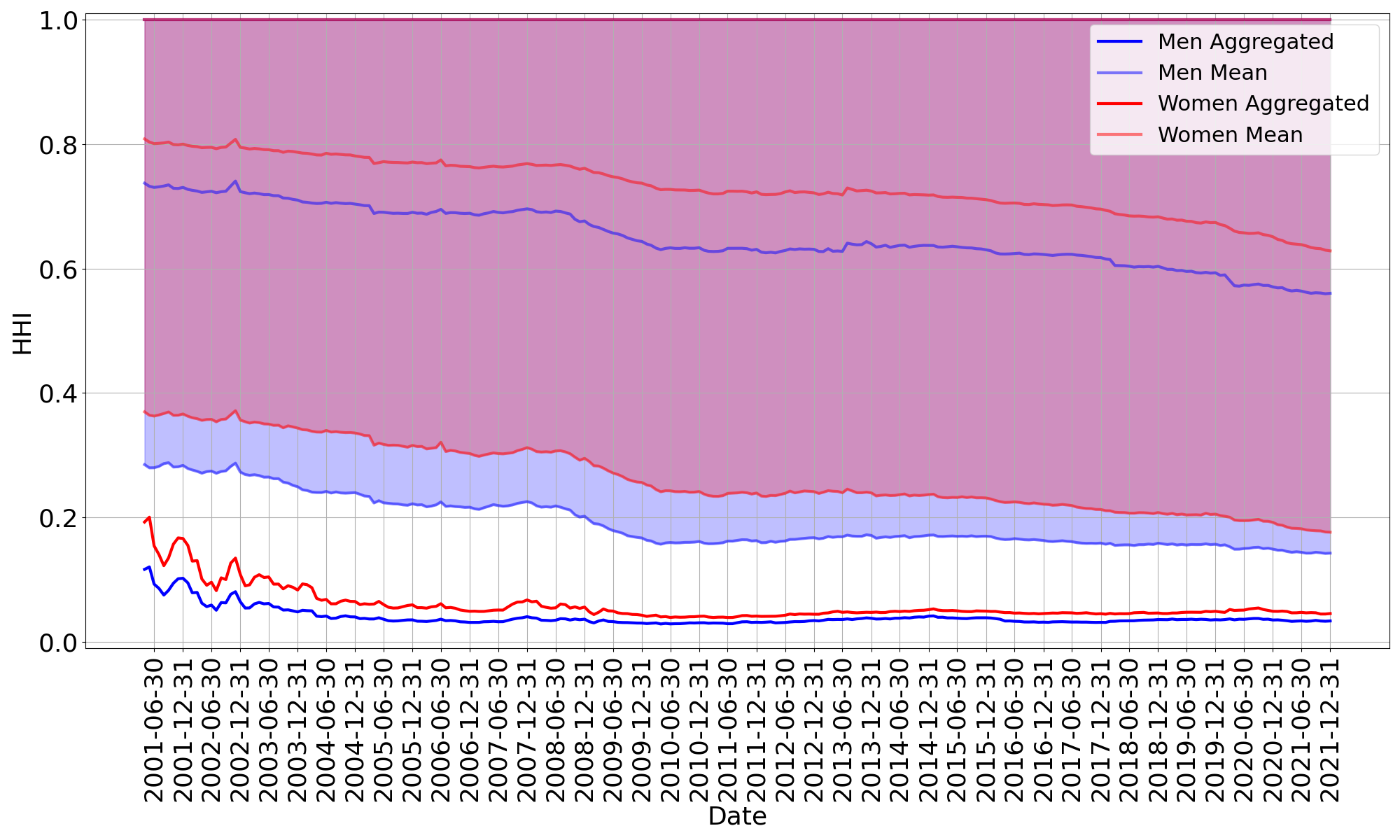}
\caption{Time evolution of HHI first and last deciles of HHI distribution for women (pink area) and men (light blue area). The red line at the center of the pink area is the mean of HHI values for women while blue line is the mean of HHI for men. For both sets of investors, the mean shows a decreasing dynamics over the years. A similar behavior is observed for the first decile. The figure also shows the HHI of the aggregated portfolios of the two sets as the red (for women)  and blue (for men) lines observed at the bottom of the figures.}\label{fig:gender}
\end{figure}

In Fig. \ref{fig:gender} we show the time evolution of the mean value of the HHI for household portfolios separately for  men and women. The figure shows the area bounded by the first and the last decile of HHI distributions as a light blue or pink shaded areas for men or women respectively. The figure also presents the time evolution of HHI of aggregated portfolios for men (blue line at the bottom of the figure) and women (red line at the bottom of the figure).  
As already evident by simultaneously analyzing  Fig. \ref{fig:hhi_dec} and Fig. \ref{fig:HHI_agg}, for household portfolios the HHI of the household aggregated portfolio  (see blue line of Fig. \ref{fig:HHI_agg}) is quite different from the mean value and the median value of the HHI distribution of household portfolios (see green and red line of Fig. \ref{fig:hhi_dec}, respectively).
This marked difference is still observed when we split the household investors in two groups characterized by the gender of investors.

This large discrepancy tells us that the high degree of concentration, i.e., high HHI value, of household portfolios does not prevent the market to build up an aggregated household portfolio that is quite diversified. This positive aggregation process benefits from two main reasons. The first reason is that although household portfolios are composed by few stocks, the selected stocks are reasonably well diversified across households. Therefore, aggregated household portfolio ends up to be less concentrated than portfolios of individual households. 
The diversification in stock selection is a spontaneous process unless coordination among households is promoted by financial marketing, by automated investment strategies or by investor coordination through social networks.  

The second aspect concerns the difference in diversification of households conditioned to the amount of portfolio value. In subsection \ref{inequality}, we show that the value of household portfolios spans up to about 10 order of magnitude with the wealth of about half of the aggregated portfolio belonging to just $1\%$ of household investors. With such pronounced heterogeneity one other possibility is that the concentration of the aggregated portfolio is primarily controlled by just the wealthier investors.

To evaluate the relative role of the value of portfolio in determining the concentration of the aggregated household portfolio we have analyzed aggregated portfolios for large set of household investors characterized by different scales of value of their individual portfolio.
Specifically, for each calendar month we have ranked all household investors from the least to the most wealthy and we have divided the set in set of men and set of women and for each of these sets we have further divided each set in 4 quartiles. For each gender and for each quartile we then have aggregated the portfolios of the selected investors into a single portfolio representative for the quartile.

\begin{figure}[h]
\centering
        \includegraphics[width=0.7\textwidth]{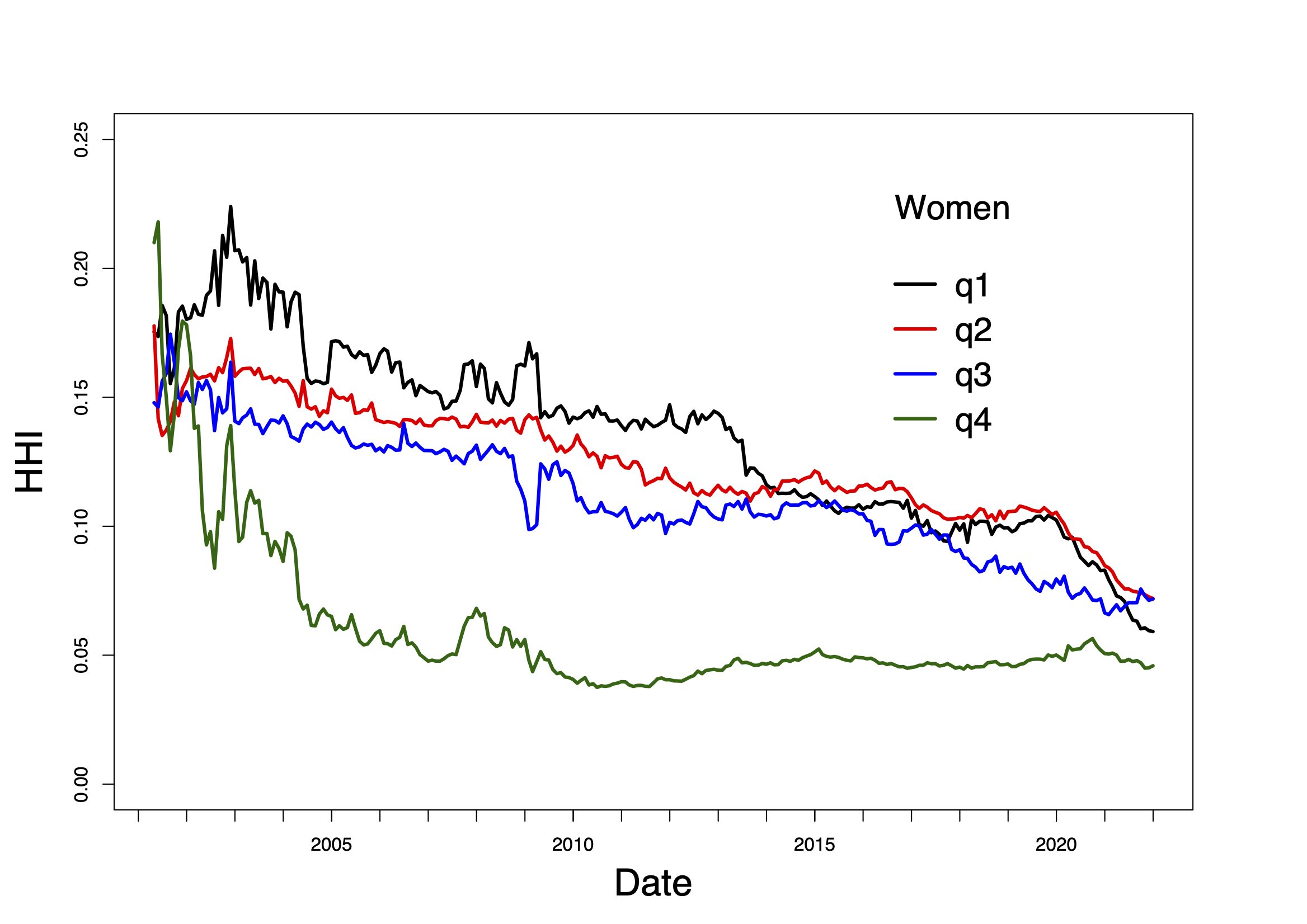}
        \includegraphics[width=0.7\textwidth]{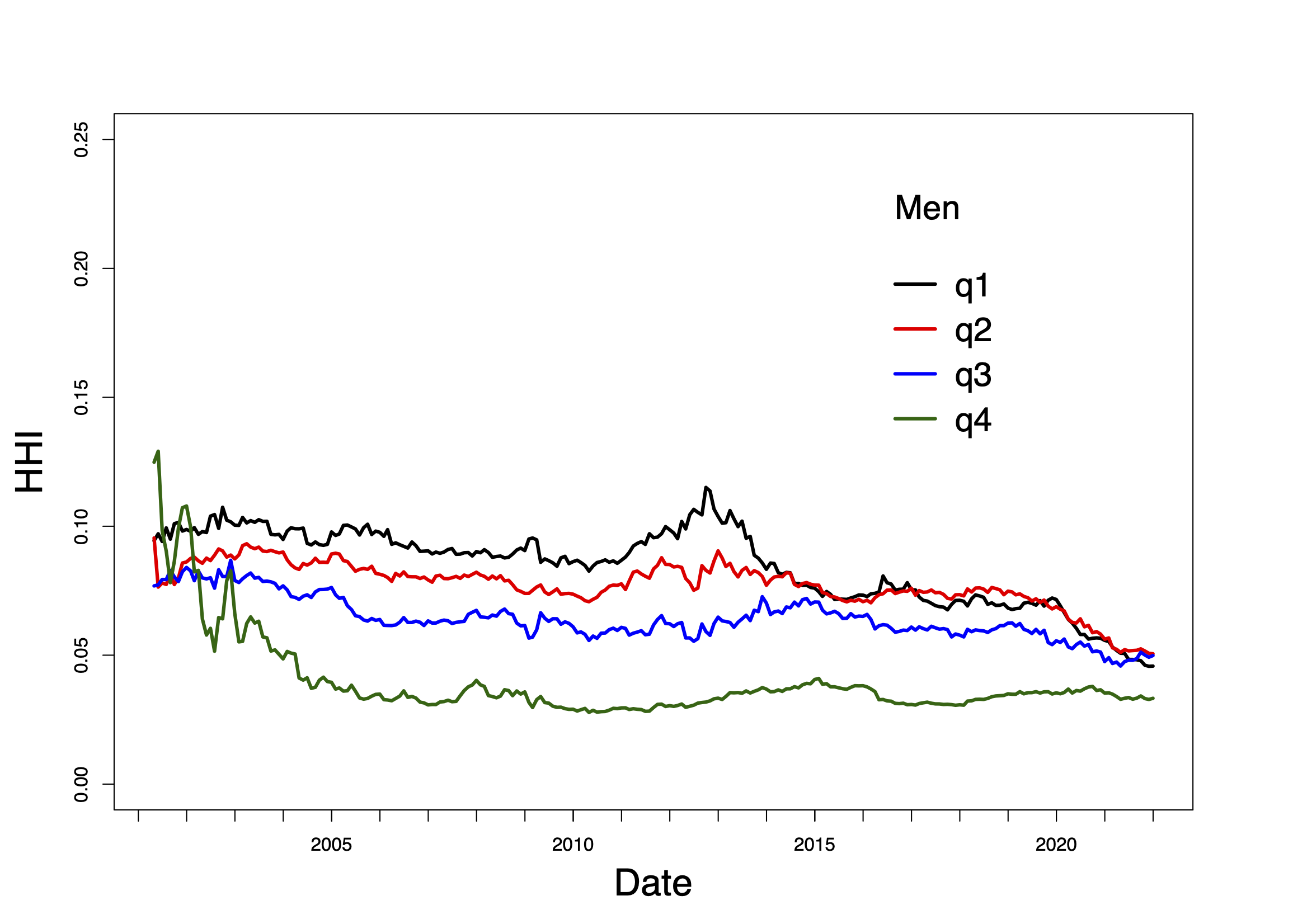}
        \caption{Time evolution of the four aggregated portfolios into whom we divide each one of  the two sets of women and men. For each month, each set has a quarter of all investors of a given gender with a long position in stocks. The first quartile $q1$ comprises investors with the lower values of the stock portfolio. The fourth quartile $q4$ comprises investors with the higher values of the stock portfolio.}
        \label{fig:gender_quart}
\end{figure}
In Fig. \ref{fig:gender_quart} we show the time evolution of the four aggregated portfolios into whom we divide the two sets of women and men. For each month, each set has a quarter of all investors of a given gender with a long position in stocks. The first quartile $q1$ comprises investors with a value of the stock portfolio ranging from a few cents of Euro to a time dependent value fluctuating around the mean value of 836 euro for women and around 1079 euro for men. The second quartile is defined by a value of the portfolio higher than the value bounding $q1$ and lower than a value fluctuating around 3488 euro for women and 4915 euro for men. The interval of the third quartile has as upper bound a value  of the portfolio higher than the value bounding $q2$ and lower than a value fluctuating around 13806 euro for women and 21103 euro for men.   

In Fig. \ref{fig:gender_quart} we notice that for both gender and for all sets of investors the HHI only for a few sets and for a few months exceed 0.2 value. For women, the HHI is on average slowly decreasing from values originally close to 0.2 down to values in the interval   
between 0.05 and 0.1. For men the HHI is on average very slowly decreasing from values close to 0.1 down to values close to 0.05. Therefore, both for women and men investors the HHI of aggregated portfolios of the eight sets of investors quantifies a degree of concentration much lower than the concentration of the portfolio of the mean (or median) investor as quantified in Fig. \ref{fig:gender}. This is true for all sets and therefore is poorly affected by the value of portfolios of investors belonging to each set.
In other words, the main driving force for the building of a not too concentrated aggregated portfolio is not the wealth of the individual investors belonging to the set, but rather the different choices in asset selection performed by the investors. A role of the wealth of the investors is also present, especially for the fourth quartile, but it is less pronounced than the one due to idiosyncratic  asset selection of the different investors.

In Fig. \ref{fig:gender_quart} we detect a difference in portfolio concentration for women and men especially for years before 2010  and for quartiles $q1$, $q2$, and $q3$. Aggregated portfolios of women are more concentrated than aggregated portfolios of men. Higher values of the HHI are observed  especially for quantiles $q1$, $q2$, and $q3$ and for years before 2010. Starting from 2010, HHI values converge  to values close to 0.05 both for women and men. 

\section{Discussion and conclusions}\label{sec5}

Our empirical results show that the degree of concentration of stock portfolios of individual Finnish investors has progressively declined both at the level of mean concentration value of portfolios of different individuals or institutions and at the level of aggregated portfolios of macro categories. 
In 2001 Finnish stock market was deeply affected by the so called dotcom bubble that international stock markets experienced during the second half of the nineties and that went burst starting from March 2000. In fact, Nokia stock was a leading company in mobile phone production and Finnish stock market at that time. Its price increased more than five times between the beginning of  1999 and April 2000 when the stock price reached its maximum value. In 1999 and 2000 Nokia capitalization was about one quarter of the entire capitalization of the Helsinki market. In other words, the entire Helsinki market was rather concentrated with Nokia being its most prominent stock. 
We perform our analysis starting from 2001 and our results show that market concentration in the Helsinki market, as quantified by the HHI, slowly decreased from about 0.4 to 0.1 in a period of time of approximately ten years. In fact, a value of the HHI close to 0.1 is observed only starting from 2011. After 2011 the market concentration present limited fluctuations around an average value close to 0.06.
The analysis of aggregated stock portfolios of macro categories of investors show that the presence of concentration during the period of time from 2001 to 2011 is primarily due to the concentration of foreign investors whereas domestic investors are less concentrated than the whole market when we consider them as a macro category of investors both for institutional and for retail Finnish investors.
This is observed in spite of the fact that individual retail investors have stock portfolios comprising a very limited number of assets. In fact, the median value of HHI for retail investors never goes below 0.5 indicating that more than half of all household portfolios have an effective number of stocks of two or less. This extreme absence of diversification at the level of the single retail investor does not imply a pronounced concentration at the level of the aggregated portfolio of all retail Finnish investors. The conclusion is rather different for foreign investors and unfortunately the data we have does not allow us to conclude whether the concentration of foreign investors is due investment choices of retail of institutional foreign investors.

By using the available metadata about Finnish retail investors we analyzed how gender affects concentration of household stock portfolios. We found that the general shape of the temporal dynamics of the HHI is rater similar between women and men investors. However a quantifiable difference is clearly detected. Over the investigated time period women have stock portfolios that are more concentrated both at the individual and at the aggregated level. However, the difference has decreased over time especially for the concentration of the aggregated portfolios.

By analysing subsets of retail investors belonging to four quartiles defined by the value of portfolio of the investor, we conclude that the low concentration of aggregated portfolios is primarily due to diversified asset selection of investors although wealthier investors present an investment pattern usually diversifying into a larger number of stocks. 

In summary, we believe our study show that data recorded for clearing and settlement has huge potential in quantifying the degree of concentration of stock portfolios owned by specific categories of investors and that this quantification can be quite informative with respect to the state of the entire market and on the role played by the different type of market participants.



\backmatter


\bmhead{Acknowledgements}

R.N. Mantegna acknowledges financial support by the MIUR PRIN Project No. 2017WZFTZP, {\it Stochastic forecasting in complex systems}.

\end{document}